\documentclass[12pt]{article}
\usepackage{color}
\usepackage{latexsym}
\usepackage{amsmath}
\usepackage{amsfonts}
\usepackage{amssymb}
\usepackage{indentfirst} 
\numberwithin{equation}{section}
\usepackage{stmaryrd}
\newcommand{\be}{\begin{equation}}
\newcommand{\ee}{\end{equation}}
 
\newcommand{\mR}{{\mathbb R}}

\title{Unitary representations of $N$-conformal Galilei group}
\author{K. Andrzejewski\thanks{e-mail: k-andrzejewski@uni.lodz.pl},
J. Gonera\\ \\
\small Department of Theoretical Physics and Computer Science, \\
\small University of \L\'od\'z,\\
\small Pomorska 149/153, 90-236 {\L}\'od\'z, Poland
}
\date{}
\begin{document}
\maketitle 
\begin{abstract}
{ All} unitary irreducible representation of centrally extended (N-odd) $N$-conformal Galilei group are constructed. The "on-shell" action of the group is derived and shown to coincide, in special but most important case, with that obtained in: J. Gomis, K. Kamimura, Phys. Rev. {\bf D85} (2012), 045023.
\end{abstract}
\section{Introduction}
The $N$-conformal Galilean algebras/groups provide the generalization of the celebrated Schr\"odinger algebra/group ($N=1$) discovered in XIX century in the context of classical mechanics \cite{b1} and heat equation \cite{b2} and subsequently rediscovered in the XX century as the maximal symmetry group  (consisting of point transformations) of free motion in quantum mechanics \cite{b3}; the mathematical structure  and  geometrical as well as physical status of the  Schr\"odinger group has been studied quite extensively \cite{b4}.
\par The higher  ($N>1$) $N$ conformal Galilean groups are also interesting, both from mathematical and physical point of view. Their detailed description was presented  in the paper Negro et. al. \cite{b5}. They were further studied in numerous papers \cite{b6}.
\par The $N$-Galilean conformal algebras split naturally into two classles: for $N$ odd they admit a one-parameter central extension  \cite{b7}-\cite{b9} while no central extension is admitted for $N$ even (except the case of two-dimensional space when the central extension exists for all  $N$).
\par One can pose the question concerning the general form of dynamics (both classical and quantum) which is invariant under $N$-conformal Galilei transformations. For $N$ odd and nontrivial central extension the answer was provided in Ref. \cite{b8}. It  appeared that the centrally extended odd $N$-conformal algebra is the symmetry algebra of free dynamics described by the Lagrangian containing $\frac{N+1}{2}$-th order time derivatives. This conclusion has been  confirmed in Ref. \cite{b10} where the orbit method \cite{b11} was applied  to the problem of classification of all invariant Hamiltonian structures. It appears that the most general canonical system  invariant under the $N$-conformal Galilei group consists  of the set of "external" canonical variables together with spin and pseudospin ones, corresponding to the $SU(2)$ and $SL(2,R)$  subgroups, respectively; the dynamics of the external variables is described by Ostrogradski Hamiltonian \cite{b12}. The authors of Ref. \cite{b8} computed also the action of $N$-conformal Galieli group on wave functions obeying the relevant Schr\"odinger  equation (on-shell action) and demonstrated its invariance.
\par In the present paper we complete the picture by finding  { all} irreducible unitary representations of centrally extended $N$-conformal Galilei group. Then we show that, when  restricted "on-shell", they yield the generalization of transformation rules derived by Gomis and Kamimura \cite{b8}. Our result  extend (to any odd $N$ ) those obtained by Perroud \cite{b13} for the case $N=1$ (Schr\"odinger group).
\section{The $N$-conformal Galilei algebra and group}
The $N$-conformal centrally extended Galilei algebra is described by the following nontrivial commutation rules:
\be
\label{e1}
\begin{split}
&[D,H]=iH,\quad [D,K]=-iK,\quad [K,H]=2iD,\\
&[J^a,J^b]=i\epsilon_{abc}J^c, [J^a,C^b_j]=i\epsilon_{abc}C^c_j,\\
&[H,C^a_j]=-ij C^a_{j-1},\quad [K,C_j^a]=i(N-j)C_{j+1}^a,\\
&[D,C^a_j]=i(\frac{N}{2}-j)C^a_j,\quad [C_j^a,C_k^b]=i\delta_{ab}\delta_{Nj+k}(-1)^{\frac{k-j+1}{2}}k!j!M;
\end{split}
\ee
here $a,b,c,\ldots=1,2,3,\, j,k,\ldots=0,1,\ldots,N$ and $N$ is odd. For $N$ even the commutation rules look the same except the last one where one should put $M=0$.
\par The structure of the algebra (\ref{e1}) is quite simple: we have three subalgebras, $su(2)$ (spanned by $J$'s), $sl(2,\mR)$ (spanned by $H$,$D$ and $K$) and the abelian one, $c_N$, which, for $N$ odd, can be centrally extended to the solvable algebra, {$\tilde c_N$}. Denoting by $g_N$ the algebra defined by the commutation rules (\ref{e1})  we have
\be
\label{e2}
g_N=(su(2)\oplus sl(2,\mR ))
\supsetplus
 (\tilde c_N).
\ee
The semidirect sum is defined by demanding that $c_N\oplus \mR$ span the representation $D^{(1,\frac N2)}\oplus D^{(0,0)}$ of $su(2)\oplus sl(2,\mR)$. 
\par The conformal Lie algebra can be easily integrated to yield the corresponding group $G_n$. We present below the form valid for any $N$ (i.e. without central extension; it is not difficult to write out the extended version).  It reads \cite{b14}
\be
\label{e3}
(g,\tilde g,{X_{ia}})*(g',\tilde g',{X'_{ia}})=(gg',\tilde g\tilde g',R_{ab}(g)X'_{ib}+X_{ja}(D^{\frac{N}{2}}(\tilde g'))^j_i);
\ee
here $g\in SU(2),\tilde g\in SL(2,\mR)$, $R(g)\in SO(3)$ is the rotation corresponding to $g$ and $D^{(\frac N2)}(\tilde g) $ is an element of $2N+1$-dimensional irreducible representation of $SL(2,\mR)$. To obtain the universal covering of $G_N$ one has only to replace $SL(2,\mR)$ by its universal covering.
\section{Irreducible unitary representations of centrally extended $N$-conformal  Galilean symmetry}
We are going to construct  unitary irreducible representations of $N$-conformal Galilean group with $N$ odd under the assumption that the central charge is nontrivial,
\be
\label{e4}
M=m{\bf 1},\quad m\neq 0.
\ee
Let us introduce new operators $\hat q_k^a,\hat p_k^a$, $a=1,2,3$, $k=0,\ldots,\frac {N-1}{2}$, by the formulae
\be
\label{e5}
\begin{split}
C^a_k=(-1)^{k-\frac{N-1}{2}}k!p_k^a,\\
C^a_{N-k}=m(N-k)!q_k^a.
\end{split}
\ee
The new operators obey the canonical commutation rules
\be
\label{e6}
[q_k^a,p_l^b]=i\delta_{ab}\delta_{kl},
\ee
and generate the Weyl group. The latter possess unique irreducible unitary representation. Once it is written out one easily constructs the unitary representation of solvable subgroup of (centrally extended ) $N$-conformal Galilei group, generated by $C_k^a$ and $M$.
\par
The remaining nontrivial commutation relations of $N$-conformal algebra, written in terms of $q$'s and $p$'s, read  
\be
\label{e7}
\begin{split}
&[J^a,q_k^b]=i\epsilon_{abc}q_k^c,\\
&[J^a,p_k^b]=i\epsilon_{abc}p_k^c,\\
&[D,q_k^a]=i(k-\frac N2)q_k^a,\\
&[D,p_k^a]=i(\frac N2-k)p_k^a,\ k= 0,\ldots ,\frac{N-1}{2},\\
&[K,q_k^a]=ik(N-k+1)q_{k-1}^a,\ k=0,\ldots,\frac{N-1}{2},\\
&[K,p_k^a]=-i(N-k)(k+1)p_{k+1}^a,\ k=0,\ldots,\frac{N-3}{2},\\
&[K,p^a_{\frac{N-1}{2}}]=im\left(\frac{N+1}{2}\right)^2q^a_{\frac{N-1}{2}},\\
&[H,q_k^a]=-iq^a_{k+1},\ k=0,\ldots,\frac{N-3}{2},\\
&[H,q^a_{\frac{N-1}{2}}]=\frac{-i}{m}p^a_{\frac{N-1}{2}},\\
&[H,p^a_k]=ip_{k-1}^a,\ k=0,\ldots,\frac{N-1}{2}.
\end{split}
\ee
The important point is that one can construct the operators belonging to the universal enveloping algebra of Heisenberg algebra which \cite{b8,b10}:
\begin{enumerate}
\item [(i)] obey the commutation rules of $su(2)\oplus sl(2,\mR)$;
\item[(ii)]obey the same commutation rules with $q$'s and $p$'s as $\vec J,H, D$ and $K$.
\end{enumerate}
 They are defined as follows
\be
\label{e8}
\begin{split}
&A=m\sum_{k=0}^{\frac{N-3}{2}}p_k^aq_{k+1}^a+\frac 12(p_{\frac{N-1}{2}})^2,\\
&B=-m\sum_{k=0}^{\frac{N-3}{2}}(k+1)(N-k)p_{k+1}^aq_k^a+\frac{m^2}{2}\left(\frac{N+1}{2}\right)^2(q^a_{\frac{N-1}{2}})^2,\\
&C=\frac{m}{2}\sum_{k=0}^{\frac{N-1}{2}}(\frac N2-k)(q_k^ap_k^a+p_k^aq_k^a),\\
&L^a=\epsilon_{abc}\sum_{k=0}^{\frac{N-1}{2}}q_k^bp_k^c.
\end{split}
\ee
It is easy to check that
\be
\label{e9}
\begin{split}
&[B,C]=imB,\\
&[C,A]=imA,\\
&[A,B]=-2miC,\\
&[L^a,L^b]=i\epsilon_{abc}L^c,
\end{split}
\ee
and that the commutation rules (\ref{e7}) are obeyed with $J^a,D,K,H$ replaced by $L^a,\frac 1m C,\frac 1mB$, and $\frac 1m A$ respectively.
Therefore the operators
\be
\label{e10}
\begin{split}
&s^a=J^a-L^a,\\
&d=D-\frac 1m C,\\
&h=H-\frac 1mA,\\
&k=K-\frac 1m B,
\end{split}
\ee
obey the $su(2)\oplus sl(2,\mR)$ algebra commutation rules and commute with $q$'s and $p$'s. Consequently, eqs. (\ref{e10}) define the decomposition of $su(2)\oplus sl(2,\mR)$ generators  into "external " and "internal "parts in full analogy with the well-known decomposition of total angular momentum into the orbital one and spin. 
This allows us to conclude that the space carrying an unitary representation of $N$-conformal Galilei group is the tensor product of  $L^2(\mR^{3\frac{(N+1)}{2}})$ which carries the representation of Weyl group generated by $q_k^a$ and $p_k^a$, $k=0,1\ldots,\frac{N-1}{2}$, $a=1,2,3$ and the carrier space of the representation of $SU(2)\times SL(2,\mR)$ group generated by $\vec s,d,h$ and $k$.
\par
The irredcible representations under consideration are classified by the eigenvalues $m$ of the central charge $M$ and the choice of the irreducible representations of $SU(2)$ and $SL(2,\mR)$  groups mentioned above.  The representations of $SU(2)$ are uniquely determined by the values of the Casimir operator 
\be
\label{e11}
\mathcal{C}_1=\hat\vec s^2,
\ee
while in the $SL(2,\mR)$ case apart from the eigenvalue of the second Casimir operator
\be\label{e12}
\mathcal{C}_2=\hat h\hat k+\hat k\hat h-2\hat d^2,
\ee
we need some additional information concerning  the spectrum of one generator \cite{b15}.
\par
Let us start with the case $\mathcal{C}_1=\mathcal{C}_{2}=0$. Then the representation of "internal" $SU(2)\times Sl(2,\mR)$ group is trivial and the total representation space is $L^2(\mR^{3\frac{(N+1)}{2}})$ consisting of functions $\psi_(\vec q_0,\ldots,\vec q_{\frac{N-1}{2}})$  square integrable with respect to the standard Lebesgue measure. Due to the identification (\ref{e5}) the action of centrally extended ableian subgroup generated by the $C_N$  subalgebra is easily expressible in terms of standard action of Weyl group.
It reads
\begin{align}
\label{e13}
&\left(e^{i\sum_{ k=0}^N\vec x_k\vec C_k}\psi	\right)(\vec q_0,\ldots,\vec q_{\frac{N-1}{2}})=e^{\frac{im}{2}\sum_{k=0}^{\frac{N-1}{2}}(-1)^{k-\frac{N-1}{2}}k!(N-k)!\vec x_k\vec x_{N-k}}\nonumber \\
&e^{im\sum_{k=0}^{\frac{N-1}{2}}(N-k)!\vec x_{N-k}\vec q_k}
\psi(\vec q_0+(-1)^{\frac{1-N}{2}}\vec x_0,\ldots,\nonumber\\
&\vec q_j+(-1)^{j+\frac{1-N}{2}}j!\vec x_j,
\ldots,\vec q_{\frac{N-1}{2}}+(\frac{N-1}{2})!\vec x_{\frac{N-1}{2}}).
\end{align}
The action of $SU(2)$ subgroup is standard 
\be\label{e14}
(U(g)\psi)(\vec q_0,\ldots,\vec q_\frac{N-1}{2})=\psi(\overrightarrow {R^{-1} (g)q_0},\ldots,\overrightarrow {R^{-1} (g)q_{\frac{N-1}{2}}}),
\ee
where $g\in SU(2)$ and $R(g)\in SO(3)$ is the corresponding rotation.
\par
It remains to find the action of $SL(2,\mR)$ subgroup. To this end we use the Iwasawa decomposition \cite{b13,b16} of $SL(2,\mR)$
\be\label{e15}
\tilde g=\left(
\begin{array}{cc}
\alpha &\beta\\
\gamma &\delta
\end{array}
\right)=
\left(
\begin{array}{cc}
\cos(\theta) &\sin(\theta)\\
-\sin(\theta) &\cos(\theta)
\end{array}
\right)
\left(
\begin{array}{cc}
u &0\\
0&u^{-1}
\end{array}
\right)
\left(
\begin{array}{cc}
1 &v\\
0 &1
\end{array}
\right),\ u>0.
\ee
By noting that the $sl(2,\mR)$ generators in defining representation can be written as 
\be\label{e16}
D=-\frac{i}{2}\sigma_3,\ K=-i\sigma_+,\ H=i\sigma_-,
\ee
we rewrite eq. (\ref{e15}) in form
\be\label{e17}
\tilde g=e^{i\theta (H+K)}e^{i\lambda D}e^{ivK},\quad u=e^{\frac{\lambda}{2}}.
\ee
It is, therefore, sufficient to determine the action of one-parameter subgroups generated by $D,K$ and ${\mathcal H}=H+K$. The relevant generators in the representation  under consideration are $\frac{1}{m}C,\frac{1}{m}B$ and $\frac{1}{m}(A+B)$, respectively.
\par
Let us start with the dilatation generator. Taking into account the explicit form of the operator $C$ we find ($U(\lambda)=e^{i\lambda D}$)
\be\label{e18}
(U(\lambda)\psi)(\vec q_0,\ldots,\vec q_{\frac{N-1}{2}})=e^{\frac{3\lambda}{16}(N+1)^2}\psi(e^{\frac{N\lambda}{2}}\vec q_0,
\ldots,e^{(\frac{N}{2}-j)\lambda}\vec q_j,\ldots,e^{\frac{\lambda}{2}}\vec q_{\frac{N-1}{2}}).
\ee
The action of $U(v)=e^{ivK}$ is slightly more complicated.  Detailed computation are given in Appendix A; we quote here only the final result
\be\label{e19}
(U(v)\psi)(\vec q_0,\ldots,\vec q_{\frac{N-1}{2}})=e^{iF(v)}\psi(\vec q_0(v),
\ldots,\vec q_{\frac{N-1}{2}}(v)),
\ee
where 
\be
\label{e19b}
\begin{split}
&\vec q_j(v)=\sum_{p=0}^j\dbinom{j}{p}\frac{(N-j+p)!}{(N-j)}(-v)^p\vec q_{j-p}\\
&F(v)= \frac{m}{2}\left(\frac{N+1}{2}\right)^2\int_0^vdu\vec q_{\frac{N-1}{2}}^2(u).
\end{split}
\ee
Finally, consider the action of compact subgroup generated by $\mathcal{H}=H+K=\frac1m(A+B)$. In this case the generator is a second order differential operator; therefore, the action of compact subgroup is nonlocal. Denoting by $\mathcal{H}(\underline{\vec q };\underline{\vec q}';\theta)$ the kernel of $\exp(i\theta\mathcal{H})$ one can write 
\be\label{e20}
\begin{split}
&(e^{i\theta\mathcal{H}}\psi)(\vec q_0,\ldots,\vec q_{\frac{N-1}{2}})=\\
&\int d^3\vec q_0'\cdots  d^3\vec q_{\frac{N-1}{2}}'\mathcal{H}(\vec q_0,\ldots,\vec q_{\frac{N-1}{2}};\vec q_0',\ldots,\vec q_{\frac{N-1}{2}}';\theta)\psi(\vec q_0',\ldots,\vec q_{\frac{N-1}{2}}').
\end{split}
\ee
The   algorithm of constructing the kernel  $\mathcal{H}(\underline{\vec q};\underline{\vec q}';\theta) $ is presented in Appendix B. Due to the fact that $\mathcal{H}$ is defined by a quadratic form in canonical variables the problem is obviously exactly solvable; the final form is, however, quite involved.
\par Let us now consider the general case $\mathcal{C}_1\neq 0$ and/or $\mathcal{C}_2\neq 0$. Due to our decomposition of generators into "external" and "internal" parts we conclude that the carrier space  consists of square integrable functions taking their values in the representation space of the representation of $SU(2)\times Sl(2,\mR)$ (or, more generally, the universal covering of $SL(2,\mR)$ generated by $\vec s,d,k$  and $h$;  note that due to noncompactness  of $SL(2,\mR)$ the relevant unitary representation is infinitedimesnional (except the case $\mathcal{C}_2=0$). The complete classification of unitary irreducible representation  of $SL(2,\mR)$ and its universal covering is well-known \cite{b15,b17}. For  convenience we use the matrix notation both for representations of $SU(2)$ and $SL(2,\mR)$ (in the latter case the matrices are infinitedimensional so  in practice it is better to work with functional realizations).  Let $D$ and $\Delta$ denote the relevant matrix representations of $SU(2)$ and  $SL(2,\mR)$, respectively. In order to write out the action of group we first note that that our relations (\ref{e13}) remain unchanged. The $SU(2)$ subgroup acts as follows
\be\label{e21}
(U(g)\psi)_{\alpha p}(\vec q_0,\ldots,\vec q_{\frac{N-1}{2}})=D_{pr}(g)\psi_{\alpha r}(\overrightarrow {R^{-1} (g)q_0},\ldots,\overrightarrow {R^{-1} (g)q_{\frac{N-1}{2}}}).
\ee
As far as $SL(2,\mR)$ is concerned we obtain 
\be\label{e22}
(U(e^{i\lambda D})\psi)_{\alpha p}(\vec q_0,\ldots,\vec q_{\frac{N-1}{2}})=e^{\frac{3\lambda}{16}(N+1)^2}\Delta_{\alpha\beta}(e^{i\lambda D})\psi_{\beta p}(e^{\frac{N}{2}}\vec q_0,
\ldots,e^{\frac{\lambda}{2}}\vec q_{\frac{N-1}{2}}),
\ee
\be\label{e23}
(U(v)\psi)_{\alpha p}(\vec q_0,\ldots,\vec q_{\frac{N-1}{2}})=e^{iF(v)}\Delta_{\alpha\beta}(e^{ivK})\psi_{\beta p}(\vec q_0(v),
\ldots,\vec q_{\frac{N-1}{2}}(v)),
\ee
and 
\begin{align}
\label{e24}
&(e^{i\theta(H+K)}\psi)_{\alpha p}(\vec q_0,\ldots,\vec q_{\frac{N-1}{2}})=
\int d^3\vec q_0'\cdots  d^3\vec q_{\frac{N-1}{2}}'\nonumber\\
&\mathcal{H}(\vec q_0,\ldots,\vec q_{\frac{N-1}{2}};\vec q_0',\ldots,\vec q_{\frac{N-1}{2}}';\theta)\Delta_{\alpha\beta}(e^{i\theta(H+K)})\psi_{\beta p}(\vec q_0',\ldots,\vec q_{\frac{N-1}{2}}'),
\end{align}
for dilations, conformal and compact transformations, respectively.
\section{The on-shell realization}
In this section we consider the action of $N$-conformal Galilei transformations on the solutions of Schr\"odinger equation. For simplicity we assume $\mathcal{C}_1=\mathcal{C}_2=0$ (the results are easily extendible to the general case) so our Hamiltonian takes the Ostrogradski form \cite{b8,b10,b12}.
\par
The action of Galilean conformal transformations is defined as follows. Given wave function at the time $t$ we translate it back to $t=0$, then act with an element of our group and finally translate the result again to time $t$ in formulae
\be\label{e25}
\tilde U(g,t)=e^{-itH}U(g)e^{itH} .
\ee
Note the identities 
\begin{subequations}
\label{e26}
\be
\label{e26a}
e^{-itH}e^{ix_k^aC_k^a}e^{itH}=e^{ix^a_k(t)C^a_k}, \quad x^a_k(t)=\sum_{j=k}^N\dbinom{j}{k}(-t)^{j-k}x^a_j,
\ee
\be
\label{e26b}
e^{-itH}U(g)e^{itH}=U(g), \quad g\in SU(2),
\ee
\be
\label{e26c}
e^{-itH}e^{i\lambda D}e^{itH}=e^{i\lambda D}e^{i(1-e^\lambda )tH},
\ee
\be
\label{e26d}
e^{-itH}e^{iv K}e^{itH}=e^{-2i\ln(1+vt)D}e^{iv(1+vt)K}e^{i(t-\frac{t}{1+vt})H},
\ee
\be
\label{e26e}
e^{-itH}e^{i\tau H}e^{itH}=e^{i\tau H}.
\ee
\end{subequations}
According to the fist formula  (\ref{e26a})  the "on-shell" action of the subgroup generated by $C_N$ is given by eq. (\ref{e13}) with $x_k^a$ replaced by $x_k^a(t)$. The action of $SU(2)$ subgroup remains unchanged. On the other hand eq. (\ref{e26c}) yields 
\be
\label{e27}
(\tilde U(e^{i\lambda D},t)\psi)(\vec q_0,\ldots,\vec q_{\frac{N-1}{2}};t)=e^{\frac{3\lambda}{16}(N+1)^2}\psi(e^{\frac{N}{2}}\vec q_0,
\ldots,e^{\frac{\lambda}{2}}\vec q_{\frac{N-1}{2}};e^\lambda t).
\ee
The action of conformal transformation is slightly more complicated. By virtue of eqs. (\ref{e18}), (\ref{e19}), (\ref{e19b}) and (\ref{e26d}) one obtains
\be
\label{e28}
\begin{split}
&(\tilde U(e^{ivK},t)\psi)(\vec q_0,\ldots,\vec q_{\frac{N-1}{2}};t)=
(1+vt)^{-\frac{3}{8}(N+1)^2}e^{iF(\frac{v}{1+vt})}\\
&\psi((1+vt)^{-N}\vec q_0\left(\frac{v}{1+vt}\right),
\ldots,(1+vt)^{-1}\vec q_{\frac{N-1}{2}}\left(\frac{v}{1+vt}\right);\frac{t}{1+vt}).
\end{split}
\ee
Finally,
\be
\label{e29}
(\tilde U(e^{i\tau H},t)\psi)(\vec q_0,\ldots,\vec q_{\frac{N-1}{2}})=\psi(\vec q_0,
\ldots,\vec q_{\frac{N-1}{2}};t-\tau).
\ee
The results obtained may be compared with those contained in Ref. \cite{b8}. We find that they are in full agreement.
\section{ Concluding remarks}
We constructed { all} unitary representations of centrally extended $N$-conformal Galilei groups. They provide the framework for constructing the "elementary"  quantum mechanical systems with $N$-conformal Galilei group as a symmetry group.
\par
{
{\bf Acknowledgments:} Special thanks are to Piotr Kosi\'nski for suggesting this topic and  useful remarks which allowed us to improve the paper. The discussions with Cezary Gonera and Pawe\l \  Ma\'slanka are gratefully acknowledged. This work  is supported  in part by  MNiSzW grant No. N202331139.}
\appendix
\section{Appendix}
We find here the action of conformal transformations. To this end let us write it in the form
\be
\label{ea1}
(e^{ivK}\psi)(\vec q_0,\ldots,\vec q_{\frac{N-1}{2}})=\psi(e^{ivK}\vec q_0e^{-ivK},
\ldots,e^{ivK}\vec q_{\frac{N-1}{2}}e^{-ivK})e^{ivK}\cdot 1.
\ee
Denoting 
\be
\label{ea2}
\vec q_j(v)=e^{ivK}\vec q_je^{-ivK},
\ee
we find
\be
\label{ea3}
\frac{d\vec q_j(v)}{dv}=ie^{ivK}[K,\vec q_j]e^{ivK}=-j(N-j+1)\vec q_{j-1}(v).
\ee
Together with the initial conditions $\vec q_j(0)=\vec q_j$ equations (\ref{ea3}) yield
\be
\label{ea4}
\vec q_j(v)=\sum_{p=0}^j\dbinom{j}{p}\frac{(N-j+p)!}{(N-j)!}(-v)^p\vec q_{j-p}.
\ee
To compute $e^{ivK}\cdot 1$ we decompose $K=\frac{1}{m}B$ as follows
\be
\label{ea5}
K=\left(i\sum_{k=0}^{\frac{N-3}{2}}(k+1)(N-k)\vec q_k\frac{\partial}{\partial \vec q_{k+1}}\right)+\left(\frac{m}{2}\left(\frac{N+1}{2}\right)^2\vec q^2_{\frac{N-1}{2}}\right)=X+Y,
\ee
and put 
\be
\label{ea6}
e^{iv(X+Y)}=Z(v)e^{ivX}.
\ee
Then $Z(v)$ obeys
\be
\label{ea7}
\frac{dZ(v)}{dv}=ie^{iv(X+Y)}Ye^{-iv(X+Y)}Z(v)=\frac{im}{2}\left(\frac{N+1}{2}\right)^2\vec q^2_{\frac{N-1}{2}}(v)Z(v),
\ee
which yields ($Z(0)=1$):
\be
\label{ea8}
Z(v)=\exp\left(\frac{im}{2}\left(\frac{N+1}{2}\right)^2\int_0^vdu\vec q_{\frac{N-1}{2}}^2(u)\right).
\ee
By virtue of (\ref{ea6}) and (\ref{ea8}) one obtains 
\be
\label{ea9}
\begin{split}
&e^{ivK}\cdot 1=\exp(iF(v)),\\
&F(v)= \frac{m}{2}\left(\frac{N+1}{2}\right)^2\int_0^vdu\vec q_{\frac{N-1}{2}}^2(u).
\end{split}
\ee
\section{Appendix}
Our aim is to describe the dynamics (both classical and quantum) generated by the "Hamiltonian" resulting from Iwasawa  decomposition (\ref{e17}). To this end we define the following Hamiltonian
\begin{align}
\label{eb1}
&\mathcal{H}=\left(\sum_{k=0}^{n-1}p_kq_{k+1}+\frac{1}{2m}p_n^2\right)+\nonumber\\
&\left(-\sum_{k=0}^{n-1}(k+1)(2n+1-k)p_{k+1}q_k+\frac{m(n+1)^2}{2}q_n^2\right),
\end{align}
where for simplicity we define $N=2n+1$ and skipped vector indices.
\par
The relevant canonical (or Heisenberg) equations of motion read
\be
\label{eb2}
\begin{split}
&\dot q_k=q_{k+1}-(2(n+1)-k)kq_k,\\
&\dot p_k=(k+1)(2n+1-k)p_{k+1}-p_{k-1},\\
&\dot q_n=\frac{1}{m}p_n-(n+2)nq_{n-1},\\
&\dot p_n=-m(n+1)^2q_n-p_{n-1}.\\
\end{split}
\ee
Eqs. (\ref{eb2}) can be solved as follows.  Define
\be
\label{eb3}
\begin{split}
&q_k=k!\xi_k,\\
&p_k=(-1)^{n-k}(2n+1-k)!m\xi_{2n+1-k},
\end{split}
\ee
for $k=0,\ldots,n$. In terms of new variables eqs. (\ref{eb3}) read (cf. Ref. \cite{b18})
\be
\label{eb4}
\dot \xi_k=(k+1)\xi_{k+1}-(2(n+1)-k)\xi_{k-1}, \quad k=0,\ldots,2n+1	.
\ee
The boundary  conditions $q_{-1}=0,p_{-1}=0$  yield $\xi_{-1}=0,\xi_{2(n+1)}=0$ which, together with (\ref{eb4}) implies $\xi_k=0$  for $k\leq -1$ or  $k\geq 2(n+1)$.  Let $x^k$ be real variable and 
\be
\label{eb5}
\xi(x,t)=\sum_{k=0}^{2n+1}\xi_{k}(t)x^k.
\ee
Eqs. (\ref{eb4}) can be summarized as follows 
\be
\label{eb6}
\frac{\partial\xi(x,t)}{\partial t}=(1+x^2)\frac{\partial \xi(x,t)}{\partial x}-Nx\xi(x,t).
\ee
By virtue of the boundary conditions for $\xi_k$ we are looking for the solutions of eq. (\ref{eb6}) which are polynomials  of degree $2n+1$ in $x$. They  read 
\be
\label{eb7}
\xi(x,t)=\sum_{\overset{l=-(2n+1)}{l\textrm {-odd}}}^{2n+1}  s_le^{ilt}P_l(x),
\ee  
\be
\label{eb8}
P_l(x)=(1+ix)^{\frac{2n+1+l}{2}}(1-ix)^\frac{2n+1-l}{2}.
\ee
Let us note  the following properties of the polynomials $P_l(x)$:
\be
\label{eb9}
\overline {P_l(x)}=P_{-l}(x),
\ee
and 
\be
\label{eb10}
\int_{-\infty}^{\infty}d\mu (x)\overline {P_l(x)}P_{l'}(x)=\delta_{ll'},\quad d\mu (x)=\frac{1}{\pi}\frac{dx}{(1+x^2)^{2(n+1)}}.
\ee
The polynomials $P_l(x)$ form an on orthonormal basis in the space of polynomials of degree $2n+1$.
\par Note that $\xi(x,t)$ is real so eq. (\ref{eb9}) implies
\be
\label{eb11}
\overline{s}_l=s_{-l}.
\ee
By comparing eqs. (\ref{eb5}) and (\ref{eb7}) one finds 
\be
\label{eb12}
\xi_k(t)=\sum_{l=-(2n+1)}^{2n+1}i^k\beta_{kl}e^{ilt}s_l,
\ee
where the coefficients  $\beta_{kl}$ are defined  through
\be
\label{eb13}
\sum_{k=0}^{2n+1}\beta_{kl}x^k=(1+x)^{\frac{2n+1-l}{2}}(1-x)^{\frac{2n+1-l}{2}}.
\ee
Note  the following properties of $\beta_{kl}$
\be
\label{eb14}
\begin{split}
&\beta_{kl}=\sum_{m=0}^k(-1)^m\dbinom{\frac{2n+1-l}{2}}{m}\dbinom{\frac{2n+1+l}{2}}{k-m},\\
&\beta_{k-l}=(-1)^k\beta_{kl},\\
&\beta_{2n+1-k,l}=(-1)^{\frac{2n+1-l}{2}}\beta_{kl}.
\end{split}
\ee
{ We will see later that the coefficient $\beta_{kl}$  have a nice interpretation in the language of finite-dimensional representations of  $SL(2,\mR)$ (see eq. (\ref{ec4})).}
\par
Eqs. (\ref{eb12}) define new dynamical variables 
\be
\label{eb15}
s_l(t)=e^{ilt}s_l,\quad l=-(2n+1),-(2n-1),\ldots, (2n-1),2n+1;
\ee
in order to compute their Poisson brackets we use (\ref{eb3}) and (\ref{eb5}) to write
\be
\label{eb16}
\xi(x,t)=\sum_{k=0}^{n}\frac{q_k(t)}{k!}x^k+\sum_{k=n+1}^{2n+1}\frac{(-1)^{n+1-k}p_{2n+1-k}(t)x^k}{mk!}.
\ee
Eq. (\ref{e16}) together with the canonical Poisson brackets yield
\be
\label{eb17}
\{\xi(x,t),\xi(y,t)\}=\frac{(-1)^{n+1}}{m(2n+1)!}(x-y)^{2n+1}.
\ee
On the other hand 
\be
\label{eb18}
\{\xi(x,t),\xi(y,t)\}=\sum_{\overset{l,l'=-(2n+1)}{l,l'\textrm{-odd}}}^{2n+1}\{s_l,s_{l'}\}e^{i(l+l')t}P_l(x)P_{l'}(y).
\ee
The left-hand side does not depend on $t$. Therefore,
\be
\label{eb19}
\{s_l,s_{l'}\}=F(n,l)\delta_{l,-l'},
\ee
and
\be
\label{eb20}
\{\xi(x,t),\xi(y,t)\}=\sum_{\overset{l,l'=-(2n+1)}{l\textrm{-odd}}}^{2n+1}F(n,l)P_l(x)P_{-l}(y).
\ee
Putting $y=0$ and using (\ref{eb17}) one arrives at the following relation
\be
\label{eb21}
\frac{(-1)^{n+1}}{m(2n+1)!}x^{2n+1}=\sum_{\overset{l=-(2n+1)}{l\textrm{-odd}}}^{2n+1}F(n,l)P_l(x).
\ee
One can find $F(n,l)$ using orthogonality relation (\ref{eb10}). Alternatively, putting $x=\tan \phi$  one gets
\be
\label{eb22}
\frac{(-1)^{n+1}}{m(2n+1)!}\sin^{2n+1}(\phi)=\sum_{\overset{l=-(2n+1)}{l\textrm{-odd}}}^{2n+1}F(n,l)e^{il\phi},
\ee
leading to 
\be\label{eb23}
F(n,l)=\frac{-i(-1)^{\frac{2n+1+l}{2}}}{m2^{2n+1}\left(\frac{2n+1+l}{2}\right)!\left(\frac{2n+1-l}{2}\right)!},
\ee
or, by virtue of (\ref{eb19}),
\be
\label{eb24}
\{s_l,s_{l'}\}=\frac{-i(-1)^{\frac{2n+1+l}{2}}}{m2^{2n+1}\left(\frac{2n+1+l}{2}\right)!\left(\frac{2n+1-l}{2}\right)!}\delta_{l,-l'}.
\ee
It remains to express the Hamiltonian (\ref{eb1}) in terms of new variables. Using (\ref{eb1}), (\ref{eb3}) and (\ref{eb12}) one obtains
\begin{align}
\label{eb25}
&\mathcal{H}=\frac{m}{2}\sum_{\overset{l,l'=-(2n+1)}{l,l'\textrm{-odd}}}^{2n+1}s_ls_{l'}e^{i(l+l')t}\Big(\sum_{k=0}^{2n}(-1)^{2n+1-k}(k+1)!(2n+1-k)!\nonumber\\
& (\beta_{2n+1-k,l}\beta_{k+1,l'}-\beta_{2n-k,l}\beta_{k,l'})\Big).
\end{align}
Using (\ref{eb14}) we find
\begin{align}
\label{eb26}
&\sum_{k=0}^{2n}(-1)^{2n+1-k}(k+1)!(2n+1-k)!
 (\beta_{2n+1-k,l}\beta_{k+1,l'}-\beta_{2n-k,l}\beta_{k,l'})\nonumber\\
&=\sum_{k=0}^{2n}(k+1)!(2n+1-k)! (\beta_{2n+1-k,l}\beta_{k+1,-l'}+\beta_{2n-k,l}\beta_{k,-l'}) \nonumber\\
&=(-1)^{\frac{2n+1-l}{2}}\sum_{k=0}^{2n}(k+1)!(2n+1-k)! (\beta_{k,l}\beta_{k+1,-l'}+\beta_{k+1,l}\beta_{k,-l'}).
\end{align}
To proceed further note that $P_l(x)$ obeys 
\be
\label{eb27}
\big((1+x^2)\frac{d}{dx}-(2n+1)x\big)P_l(x)=ilP_l(x).
\ee
Inserting the expansion
\be
\label{eb28}
P_l(x)=\sum_{k=0}^{2n+1}i^k\beta_{kl}x^k,
\ee
one derives the recurrence relation 
\be
\label{eb29}
\begin{split}
(k+1)\beta_{k+1,l}+(2(n+1)-k)\beta_{k-1,l},-l\beta_{k,l}&=0,\quad k=1,\ldots,2n+1\\
\beta_{1l}-l\beta_{0l}&=1.
\end{split}
\ee
It easy to check with the help of (\ref{eb29}) that
\begin{align}
\label{eb30}
&(l-l')\sum_{k=0}^{2n+1}k!(2n+1-k)!\beta_{k,l}\beta_{k,-l'}\nonumber\\
&=2\sum_{k=0}^{2n}(k+1)!(2n+1-k)!(\beta_{k+1,l}\beta_{k,-l'}+\beta_{kl}\beta_{k+1,-l'}).
\end{align}
We will show that
\be
\label{eb31}
\sum_{k=0}^{2n+1}k!(2n+1-k)!\beta_{k,l}\beta_{k,l'}=G(n,l)\delta_{ll'},
\ee
where
\be
\label{eb32}
G(n,l)=2^{2n+1}\left(\frac{2n+1+l}{2}\right)!\left(\frac{2n+1-l}{2}\right)!.
\ee
{
We prove (\ref{eb31}) and (\ref{eb32}) in two ways. In the first approach we  note that the matrix $i^k\beta_{kl}$ is invertible because it relates two bases.} Putting 
\be
\label{eb33}
x^k=\sum_{\overset{l=-(2n+1)}{l\textrm{ - odd}}}^{2n+1}\gamma_{lk}P_l(x),
\ee
one gets
\be
\label{eb34}
\sum_{\overset{l=-(2n+1)}{l\textrm{ - odd}}}^{2n+1}i^k\beta_{kl}\gamma_{lk'}=\delta_{kk'},\quad \sum_{k=0}^{2n+1}i^k\gamma_{lk}\beta_{kl'}=\delta_{ll'}.
\ee
Eq. (\ref{eb31}) is equivalent to 
\be
\label{eb35}
k!(2n+1-k)!\beta_{kl}=G(n,l)\gamma_{lk}(-i)^k.
\ee
We prove (\ref{eb35}) by deriving the recurrence relation for $\gamma_{lk}$. To this end note that (\ref{eb33}) implies 
\be
\label{eb36}
\gamma_{lk}=\int_{\infty}^{\infty}d\mu(x)x^k\overline{P_l(x)}.
\ee
The operator $(1+x^2)\frac{d}{dx}-(2n+1)x$ is antyhermitean with respect to the product defined by $d\mu(x)$. Therefore,
\be
\label{eb38}
\begin{split}
&(2n+1-k)\gamma_{l,k+1}-k\gamma_{l,k-1}-il\gamma_{lk}=0,\\
& (2n+1)\gamma_{l1}-il\gamma_{l0}=0.
\end{split}
\ee
Now, $\frac{i^kk!(2n+1-k)!\beta_{kl}}{G(n,l)}$ obey the same recurrence. To find $G(n,l)$ we note that $\beta_{0l}=1$ and
\be
\label{eb39}
\gamma_{l,0}=\frac{1}{\pi}\int_{-\frac\pi 2}^{\frac\pi 2}d\phi\cos^{2n+1}(\phi) e^{-il\phi}=\frac{(2n+1)!}{2^{2n+1}(\frac{2n+1-l}{2})!(\frac{2n+1+l}{2})!},
\ee
which concludes the proof of (\ref{eb35}).
\par
{
The second proof  of eqs. (\ref{eb31}) and (\ref{eb32})  make use of the properties of finite dimensional irreducible representations of $SL(2,\mR)$. In fact, finite-dimensional irreducible representations of $SL(2,\mR)$  are classified by natural number $N$ and they matrices are of the form (see, e.g., \cite{b14}) 
\be
\label{ec1}
\Phi(A)_{mm'}=\sum_{\max(0,m'-m)}^{\min{(N-m,m')}}\dbinom{N-m}{k}\dbinom{m}{m-k}a^{N-m-k}b^kc^{m-m'+k}d^{m'-k}, 
\ee
\label{ec2}
where $  A=\left(\begin{array}{cc}a&b\\c&d\end{array}\right)\in SL(2,\mR)$ and $m,m'=0,\ldots,N$; in our case $N=2n+1$. Now let $A$ be of the form 
\be
\label{ec3}
A=\frac{1}{\sqrt{2}}\left(\begin{array}{cc}1&-1\\1&1\end{array}\right)\in SL(2,\mR) .
\ee
It is not difficult to show that 
\be
\label{ec4}
\Phi(A)_{mm'}=\frac{1}{\sqrt{2^{2n+1}}}\beta_{m',-(2n+1)+2m}.
\ee
Since  $A^TA=AA^T=Id$  one finds
\be 
\label{ec5}
\Phi(A)\Phi(A^T)=\Phi(A^TA)=Id.
\ee
Moreover,  let us note that in our case we have 
\be 
\label{ec6}
\Phi(A)_{mm'}=(-1)^{m'-m}\frac{\dbinom{2n+1}{m'}}{\dbinom{2n+1}{m}}(\Phi(A))_{m'm},
\ee 
and 
\be
\label{ec7}
\Phi(A^T)_{mm'}=(-1)^{m-m'}\Phi(A )_{mm'}.
\ee
Inserting (\ref{ec7}) into eq. (\ref{ec5}) and using eqs. (\ref{ec4}) and (\ref{ec6})  we obtain the relations  (\ref{eb31}) and (\ref{eb32}).
}
\par
Collecting all formulae we find
\be
\label{ec40}
\mathcal{H}=\frac{m}{2}\sum_{\overset{l=1}{l{\textrm  -odd}}}^{2n+1}(-1)^{\frac{2n+1-l}{2}}l2^{2n+1}
\left(\frac{2n+1-l}{2}\right)!\left(\frac{2n+1+l}{2}\right)!(s_l\overline{s}_l+\overline{s}_ls_l);
\ee
upon defining ($l=1,\ldots,2n+1$, $l$-odd)
\be
\label{eb41}
a_l =\left\{
\begin{array}{c}
2^{\frac{2n+1}{2}}\sqrt{m}\sqrt{\left(\frac{2n+1-l}{2}\right)!\left(\frac{2n+1+l}{2}\right)!} s_l, \quad \frac{2n+1+l}{2}\textrm{ - even}\\
2^{\frac{2n+1}{2}}\sqrt{m}\sqrt{\left(\frac{2n+1-l}{2}\right)!\left(\frac{2n+1+l}{2}\right)!}\overline {s}_l, \quad \frac{2n+1+l}{2}\textrm{ - odd,}
\end{array}
\right.
\ee
one arrives finally at the following result
\be
\label{eb42}
\begin{split}
&\mathcal{H}=\frac{1}{2}\sum_{\overset{l=0}{l\textrm{-odd}}}^{2n+1}(-1)^\frac{2n+1-l}{2}l(a_l\overline{a}_l+\overline{a}_la_l),\\
&\{a_l,\overline{a}_{l'}\}=-i\delta_{ll'},\quad l,l'=1,3,\ldots,2n+1.
\end{split}
\ee
Quantization yields 
\be
\label{eb43}
\begin{split}
&\mathcal{H}=\frac{1}{2}\sum_{\overset{l=0}{l\textrm{-odd}}}^{2n+1}(-1)^\frac{2n+1-l}{2}l(a_l^+{a}_l+\frac12),\\
&[a_l,{a}^+_{l'}]=\delta_{ll'},\quad l,l'=1,3,\ldots,2n+1.
\end{split}
\ee
Define further, for $k=0,1,\ldots,n$, $l=1,3,\ldots,2n+1$,
\be
\label{eb44}
\rho_{kl}=\left\{
\begin{array}{c}
\frac{i^kk!\beta_{kl}}{2^{\frac{2n+1}{2}}\sqrt{m}\sqrt{(\frac{2n+1-l}{2})!(\frac{2n+1+l}{2})!}}, \quad \frac{2n+1+l}{2}\textrm{ - even}\\
\frac{(-i)^kk!\beta_{kl}}{2^{\frac{2n+1}{2}}\sqrt{m}\sqrt{(\frac{2n+1-l}{2})!(\frac{2n+1+l}{2})!}}, \quad \frac{2n+1+l}{2}\textrm{ - odd,}
\end{array}
\right.
\ee
then
\be
\label{eb45}
q_k=\sum_{\overset{l=1}{l\textrm{-odd}}}^{2n+1}\rho_{kl}a_l+\sum_{\overset{l=1}{l\textrm{-odd}}}^{2n+1}\overline{\rho}_{kl}a_l^+,\quad k=0,\ldots,n.
\ee
The coordinate operators commute which implies
\be
\label{eb46}
\rho\overline{\rho}^T=\overline{\rho}\rho^T,
\ee
so ${\rho}\overline{\rho}^T$ is real symmetric (eq.  (\ref{eb46}) can be checked directly  using  (\ref{eb31}))  hence diagonalizable by real orthogonal transformation. Assume $\det \rho=0$; then $\det({\rho}\overline{\rho}^T)=0$ and there exists real nonzero vector $u$ such that $u^T\rho=0$.  Taking complex  conjugate  one finds $u^T\overline {\rho}=0$. Therefore, by  virtue of eq.  (\ref{eb45}) 
\be
\label{eb47}
\sum_{k=0}^nu_kq_k=0,
\ee
which  contradicts the canonical commutation rules. So we conclude that  $\rho$ is invertible.
\par 
Finally, define for $k=0,\ldots,n$,\quad $l=1,3,\ldots,2n+1$
\be
\label{eb48}
\tau_{kl}=\left\{
\begin{array}{c}
\frac{(-1)^{n-k}(2n+1-k)!\sqrt{m} i^{2n+1-k}\beta_{2n+1-k,l}}{2^{\frac{2n+1}{2}}\sqrt{(\frac{2n+1-l}{2})!(\frac{2n+1+l}{2})!}}, \quad \frac{2n+1+l}{2}\textrm{ - even}\\
\frac{(-1)^{n-k}(2n+1-k)!\sqrt{m} (-i)^{2n+1-k}\beta_{2n+1-k,l}}{2^{\frac{2n+1}{2}}\sqrt{(\frac{2n+1-l}{2})!(\frac{2n+1+l}{2})!}}, \quad \frac{2n+1+l}{2}\textrm{ - odd.}
\end{array}
\right.
\ee
Then, for  $k=0,\ldots,n$ 
\be
\label{eb49}
p_k=\sum_{l=1}^{2n+1}\tau_{kl}a_l+\sum_{l=1}^{2n+1}\overline{\tau}_{kl}a_l^+.
\ee
Note that $\rho_{kl}=\overline{\rho_{kl}}$ for $k$ even while 
$\tau_{kl}=\overline{\tau}_{kl}$ for $k$ odd.
\par
Let us introduce new canonical variables 
\be
\label{eb50}
Q_l=\frac{1}{\sqrt{2l}}(a_l+a_l^+),\quad P_l=i{\sqrt{\frac l2}}(-a_l+a_l^+).
\ee
Then the Hamiltonian takes the form
\be
\label{eb51}
\mathcal{H}=\sum_{\overset{l=0}{l\textrm{-odd}}}^{2n+1}(-1)^{\frac{2n+1-l}{2}}(\frac{P_l^2}{2}+\frac{l^2Q_l^2}{2}).
\ee
Moreover
\be
\label{eb52}
\begin{split}
&q_k=\sum_{\overset{l=0}{l\textrm{-odd}}}^{2n+1}\sqrt{2l}\rho_{kl}Q_l,\quad k\textrm{ - even};\quad
q_k=\sum_{\overset{l=0}{l\textrm{-odd}}}^{2n+1}i\sqrt{\frac2l}\rho_{kl}P_l,\quad k\textrm{ - odd}\\
&p_k=\sum_{\overset{l=0}{l\textrm{-odd}}}^{2n+1}i\sqrt{\frac2l}\tau_{kl}P_l,\quad k\textrm{ - even};\quad
p_k=\sum_{\overset{l=0}{l\textrm{-odd}}}^{2n+1}\sqrt{2l}\tau_{kl}Q_l,\quad k\textrm{ - odd}.
\end{split}
\ee
The above formulae allow us to give a simple prescription for computing the kernel  $\mathcal{H}(\vec{\underline{ q}},\vec{\underline{ q}}';\theta)$ of $\exp(i\theta\mathcal{H})$. Namely,  the canonical variables $(Q_l,P_l) $ diagonalize the Hamiltonian. Therefore, in terms of them the propagator kernel is simply the product of single propagators for harmonic oscillators; to account for the sign on the right-hand side of eq.  (\ref{eb51})   in every second term in the product the replacement 
$\theta \rightarrow -\theta $ should be made. Next, note that   (\ref{eb52})  can be viewed as the composition of the point transformation 
\be
\label{eb53}
\tilde q_k=\left\{
\begin{array}{c}
\sum_{\overset{l=1}{l\textrm{-odd}}}^{2n+1}\sqrt{2l}\rho_{kl}Q_l,\quad k\textrm{ - even}\\
\sum_{\overset{l=1}{l\textrm{-odd}}}^{2n+1}\sqrt{2l}\tau_{kl}Q_l,\quad k\textrm{ - odd,}
\end{array}
\right.
\ee
\be
\label{eb54}
\tilde p_k=\left\{
\begin{array}{c}
\sum_{\overset{l=1}{l\textrm{-odd}}}^{2n+1}i\sqrt{\frac2l}\tau_{kl}P_l,\quad k\textrm{ - even}\\
-\sum_{\overset{l=1}{l\textrm{-odd}}}^{2n+1}i\sqrt{\frac2l}\rho_{kl}P_l,\quad k\textrm{ - odd,}
\end{array}
\right.
\ee
with  a simple canonical transformation 
\be
\label{eb55}
\begin{split}
&q_k=\tilde q_k,\quad \textrm{k - even},\\
&q_k=-\tilde p_k,\textrm{\quad k - odd},\\
&p_k=\tilde p_k,\quad \textrm{k - even},\\
&p_k=\tilde q_k,\quad \textrm{k - odd},\\
\end{split}
\ee
 (\ref{eb53})  and  (\ref{eb54})  result in expressing the initial kernel in terms of new variables $\tilde q_k$ and multiplying it by an appropriate constant factor according to the formula
\be
\label{eb56}
<\underline{\tilde q}|\underline Q>=|\det(\frac{\partial f_k}{\partial Q_m})|^{-\frac12}\delta(\underline{\tilde q}-\underline{f}(\underline{Q})).
\ee
The final step is to perform the canonical transformation  (\ref{eb55}) . The corresponding kernel for it reads 
\be
\label{eb57}
<\underline{ q}|\underline{\tilde q}>=\Pi_{k\textrm{ - even}}\delta(q_k-\tilde q_k)\Pi_{k\textrm{ - odd}}\frac{1}{\sqrt{2\pi}}e^{iq
_k\tilde q_k},
\ee
i.e., we perform, the Fourier transform with respect to odd variables.
\par
Finally, let remind that up to now we have skipped the vector indices. However, the dynamics is diagonal with respect to them so one has only to multiply kernels for the propagation for all separate components.

\end{document}